\documentclass[sn-mathphys,Numbered]{sn-jnl}% Math and Physical Sciences Reference Style
\usepackage{subcaption}
\usepackage{graphicx}%
\usepackage{multirow}%
\usepackage{amsmath,amssymb,amsfonts}%
\usepackage{amsthm}%
\usepackage{mathrsfs}%
\usepackage[title]{appendix}%
\usepackage{xcolor}%
\usepackage{textcomp}%
\usepackage{manyfoot}%
\usepackage{booktabs}%
\usepackage{algorithm}%
\usepackage{algorithmicx}%
\usepackage{algpseudocode}%
\usepackage{listings}%
\usepackage{url}
\usepackage{graphicx,epstopdf}
\usepackage{pgfplots}
\usepackage{verbatim}
\theoremstyle{thmstyleone}%
\newtheorem{theorem}{Theorem}%  meant for continuous numbers
\newtheorem{proposition}[theorem]{Proposition}% 

\theoremstyle{thmstyletwo}%

\theoremstyle{thmstylethree}%

\begin{document}

\title[Article Title]{An efficient approach for searching three-body periodic orbits passing through Eulerian configuration}

%% FamilyName	-> \sur{Ploeg}f

\author*{Ivan Hristov$^{1}$,  Radoslava Hristova$^{1}$}

\affil{$^{1}$ Faculty of Mathematics and Informatics, Sofia University, 5 James Bourchier blvd.,
Sofia 1136, Bulgaria}

\email{ivanh@fmi.uni-sofia.bg}  \email{radoslava@fmi.uni-sofia.bg}


\abstract{A new efficient approach for searching three-body periodic equal-mass 
collisionless orbits passing through Eulerian configuration is presented.
The approach is based on a symmetry property of the solutions at the half period. Depending on   two  previously established symmetry types 
%small change-RM
 on the shape sphere, each solution is presented by one or two distinct initial conditions (one or two points in the search domain). A numerical search based on  Newton's method on a relatively coarse search grid for solutions with relatively small scale-invariant periods $T^{*}<70$ is conducted. The linear systems at each Newton's iteration are computed by  high order high precision Taylor series method. The search produced 12,431 initial conditions (i.c.s) corresponding to 6,333 distinct solutions. In addition to passing through
 the
 %small change-RM
  Eulerian configuration, 35 of the solutions are also free-fall ones. Although most of the found solutions are new, all linearly stable solutions among them (only 7) are old ones. Particular attention is paid to the details of the high precision computations and the analysis of accuracy. All i.c.s are given with 100 correct digits.}




\maketitle

%\begin{comment}

\section{Introduction}\label{s:Introduction}

The fundamental works of \v Suvakov and Dmitra\v sinovi\' c \cite{Suvakov:2013, Suvakov:2014} paved the way for further numerical searches and discoveries
of a large number of new three-body periodic orbits after 2013. They formulated the main principles of the searching procedure, analyzed the initial configurations, explained and applied the topological classification of the orbits, established symmetries of the solutions on the shape sphere \cite{Montgomery:shape} and also algebraic exchange symmetries of their free group elements \cite{Montgomery:braid}. They discovered many new stable periodic orbits. These papers inspired many people to repeat their
%small change-RM
 results  and  conduct their own searches. The algorithms in \cite{Suvakov:2013, Suvakov:2014} were performed in the standard double precision arithmetic and hence they have the advantage of 
  successfully capturing
  %small change-RM
   the important stable solutions without much work. However,  
   %small change-RM
   double precision computations lack the ability to find some
    %small change-RM
    unstable periodic orbits, because these orbits can by very sensitive on the initial conditions (with large Lyaponov exponents). Standing on the papers \cite{Suvakov:2013, Suvakov:2014}, Li and Liao  took a further  step in 2017  \cite{Liao:2017} when 
    they
     %small change-RM
     announced   a large number of new periodic orbits obtained from a numerical search with Newton's method. High precision floating point arithmetic combined with high order Taylor series method \cite{Jorba, Barrio:2006, Barrio:2011, Izzo}  were used for computing the linear systems in \cite{Liao:2017} in order to overcome the sensitive dependence on the initial conditions and hence to 
     % %small change-RM  `brake': stop
   overcome
    the limitations of using the standard arithmetic. 

Many of the numerical searches in the past decade, including  those  already cited  \cite{Suvakov:2013, Suvakov:2014, Liao:2017}, and others \cite{Suvakov:2014b, Suvakov:2016, Dmitrasinovic:2018, Liao:2018, Hristov:AMI22, Hristov:Bor22}, have been devoted to searching periodic orbits which pass  through the symmetrical collinear (Eulerian) configuration. More precisely, this configuration is taken as the initial one at $t=0$. As mentioned above, symmetries on the shape sphere were established in \cite{Suvakov:2013}. The orbits in \cite{Suvakov:2013} were divided into two types: ``(I) those with reflection symmetries about two orthogonal axes on the shape sphere - the equator and the zeroth meridian and (II) those with a central reflection symmetry about a single point - the intersection of the equator and zeroth meridian''.
Although these symmetries suggest some special arrangement of the bodies at $t=T/2$, where $T$ is the period of the periodic orbit, we did not find
a clear statement about this arrangement in the literature.  We analyzed all solutions from \cite{Suvakov:2013, Suvakov:2014, Liao:2017, Hristov:AMI22, Hristov:Bor22} and found that for all of them the bodies at $t=T/2$ are again in Eulerian configuration. After a correspondence with prof. Richard Montgomery, the property of the periodic orbits to pass again through Eulerian configuration at $t=T/2$ was soon proved by him. We give this proof in an appendix of this paper. We additionally observed that for solutions of symmetry type (I) the distance between bodies at $t=T/2$ is the same to those at $t=0$ and the Eulerian configuration at $t=T/2$ is essentially the same to those at $t=0$. For symmetry type (II) solutions however, the distance between bodies at $t=T/2$ is different and the Eulerian configuration is also different. The latter means that the orbits of type (II) are presented by two different initial condition' points in the search domain, while those of type (I) - by one point. This more precise statement about the configuration at $t=T/2$ needs of course an additional proof, which we do not give here. This is not however a restriction for us to compute easily the second i.c. for a solution of type (II) (if it is missed) by rescaling \cite{Suvakov:2013, Suvakov:2014} the positions and velocities at $t=T/2$ and testing the i.c. for convergence after that.

The main goal of this paper is to conduct a numerical search based on the above ``half period'' property. The proof of the property actually provides the theoretical foundation of the proposed approach. Instead of looking for orbits that satisfy the standard periodicity condition, we look for orbits that pass again through Eulerian configuration at some later time. This approach turned out to be much more efficient than the standard one, because (as numerous experiments show) Newton's method has a much larger domain of convergence. With a relatively coarse search grid and a search for solutions with relatively short scale-invariant periods $T^*=T|E|^{3/2} < 70$ (E is the energy of the orbit), we were able to find more than 12,000 initial conditions, most of them new ones. Let us mention that all the found new orbits turned out to be not stable. This is not unexpected, because stable periodic orbits are much easier to be found numerically and many of them (especially those with small invariant periods) are already found in the previous searches. Nevertheless such kind of very efficient searches can be important for further theoretical investigations of the chaotic behaviour of the three-body problem \cite{Cvitanovic:1987, Cvitanovic:1988}.

Usually the scanning of the search domain is done 
by simulation at each grid-point up to some pre-defined value of the time - the upper bound of the periods.
As said above, instead of this usual (uniform) constraint for the simulation time, in this paper we use a scale-invariant one.
This means that the upper bound of the scale-invariant period is pre-defined instead. 
The simulation time is not uniform, but different for the different points in the search domain and is determined by the energy $E$ of the point. 
In this case we have a singularity, when approaching the boundary curve of the domain (the $E=0$ curve), meaning that  
the absolute periods must grow without limit when approaching this curve.
The replacement of the usual uniform constrain with the scale invariant one, 
allows us to check the hypothesis for the existence of periodic orbits when closely approaching the boundary. 
The numerical results suggest the existence of a region in the search domain, where we have a sequence of periodic orbits
with ever increasing periods, when approaching the $E=0$ curve.

The paper consists of 5 sections.
After the present Introduction, in Sect. \ref{s:Mathematical model} we present the mathematical model (the differential equations describing the three-body motion and the initial Euler configuration) and also the used proximity functions. In Sect. \ref{s:Numerical methods} we present the used numerical methods, particularly we explain the new searching approach. 
In Sect. \ref{s:Numerical results} we discuss the parameters of simulations and their accuracy and present the numerical results.
In Sect. \ref{s:Conclusions} we summarize and draw conclusions.

\section{Mathematical model and proximity functions}
\label{s:Mathematical model}
\subsection{Differential equations}
\label{ss:Differential equations}

The differential equations describing the motion of the three-body system are derived
from Newton's second law and Newton's law of gravity:
$$
m_i\ddot{r}_i=\sum_{j=1,j\neq i}^{3}G m_i m_j \frac{(r_j-r_i)}{{\|r_i -r_j\|}^3}, i=1,2,3.
$$
where $G$ is the gravitational constant, $m_i$ are the masses and $r_i$ are the vectors
of the positions. The model treats the bodies as point masses.
Dividing the equations by $m_i$  and considering $G=1$, we have:
\begin{equation}
\label{equasecond}
\ddot{r}_i=\sum_{j=1,j\neq i}^{3}m_j \frac{(r_j-r_i)}{{\|r_i -r_j\|}^3}, i=1,2,3.
\end{equation}
In this paper we consider planar motion and equal masses of the bodies ($m_1=m_2=m_3=1$).
Hence the vectors $r_i$, $\dot{r}_i$ have two components: $r_i=(x_i, y_i)$,
$\dot{r}_i=(\dot{x}_i, \dot{y}_i).$
The following dependent variables ${vx}_i$ and ${vy}_i$ are introduced, so that ${vx}_i=\dot{x}_i, {vy}_i=\dot{y}_i.$
Then the second order system (\ref{equasecond}) can be written as a first order one:
\begin{equation}
\label{equations}
\dot{x}_i={vx}_i, \hspace{0.1 cm} \dot{y}_i={vy}_i,  \hspace{0.1 cm} \dot{vx}_i=\sum_{j=1,j\neq i}^{3}\frac{(x_j-x_i)}{{\|r_i -r_j\|}^3},  \hspace{0.1 cm} \dot{vy}_i=\sum_{j=1,j\neq i}^{3}\frac{(y_j-y_i)}{{\|r_i -r_j\|}^3}, \hspace{0.1 cm} i=1,2,3
\end{equation}
The system is solved numerically in this first order form. We have a vector of 12 unknown functions
$ X(t)={(x_1, y_1, x_2, y_2, x_3, y_3, {vx}_1, {vy}_1, {vx}_2, {vy}_2, {vx}_3, {vy}_3)}^\top$.

\begin{figure}
\centerline{\includegraphics[width=0.5\columnwidth,,keepaspectratio]{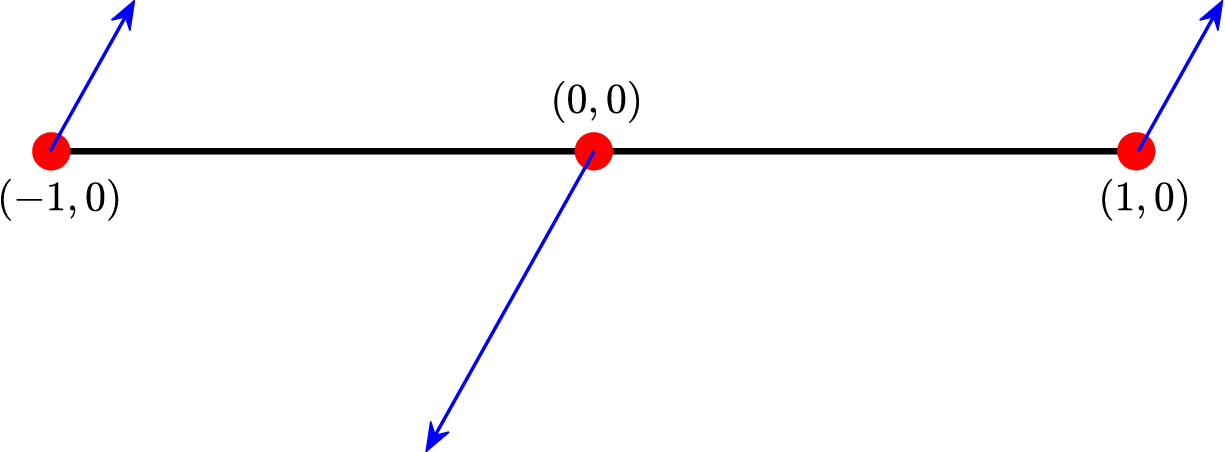}}
\caption{\footnotesize{The symmetrical collinear configuration with parallel velocities (Eulerian configuration)}}
\label{fig:euler}
\end{figure}

\subsection{Initial conditions: Euler symmetries}
\label{ss:Initial configuration}

\begin{figure}
\centerline{\includegraphics[width=0.5\columnwidth,,keepaspectratio]{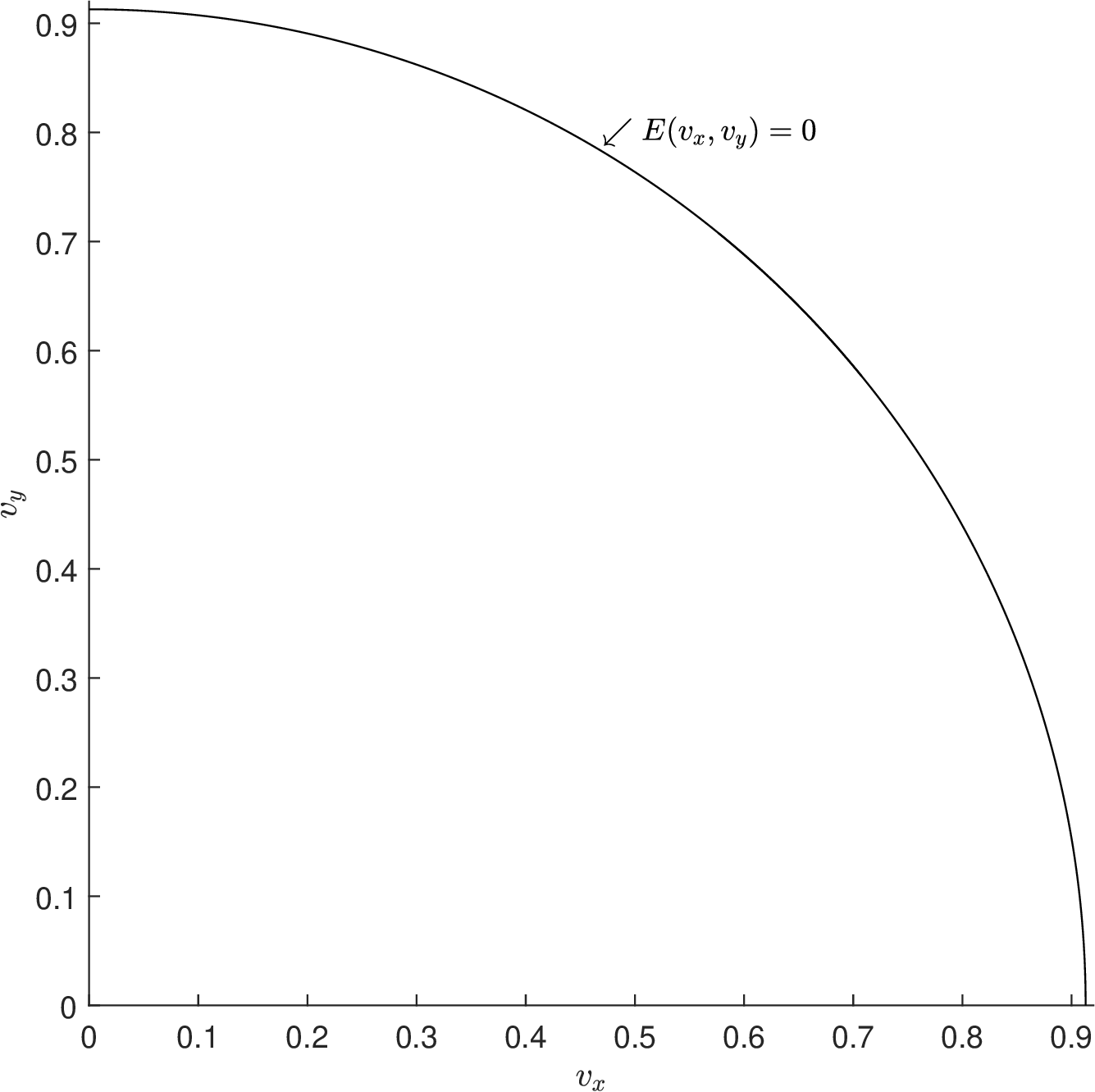}}
\caption{The 2D search domain, bounded by $v_x=0$ and $v_y=0$ axis and $E=0$ curve}
\label{fig:search domain}
\end{figure}

We consider as an initial configuration (initial positions and velocities) the symmetrical collinear with parallel velocities 
(Eulerian) configuration with two parameters $v_x>0$, $v_y>0$ \cite{Suvakov:2013, Suvakov:2014} (see Fig. \ref{fig:euler}):
\begin{center}
\begin{equation}
\label{initcond}
\begin{split}
(x_1(0),y_1(0))=(-1,0), \hspace{0.2 cm} (x_2(0),y_2(0))=(1,0), \hspace{0.2 cm} (x_3(0),y_3(0))=(0,0) \hspace{0.5 cm} \\
({vx}_1(0),{vy}_1(0))=({vx}_2(0),{vy}_2(0))=(v_x,v_y)  \hspace{2.5 cm}\\
({vx}_3(0),{vy}_3(0))=-2({vx}_1(0),{vy}_1(0))=(-2v_x, -2v_y) \hspace{1.8 cm}
\end{split}
\end{equation}
\end{center}
Here we choose a numeration with the third body in the middle.
%rmont: big addition here, regarding ``3rd condition' in email of 4/22 exchange
These initial conditions, and hence solutions, have  zero linear and zero angular momentum and have energy $E=-2.5+3({v_x}^2+ {v_x}^2).$
For initial positions $x_a (0), y_a (0)$ fixed as above, the two-dimensional space of velocities described
by equation (\ref{initcond}) is precisely the space of  all   velocities for which
the linear and angular momentum are zero and  for which the moment of inertia or   `size' 
variable $I = {x_1}^2 + {y_1}^2 +  {x_2}^2 + {y_2}^2 + {x_3}^2 + {y_3}^2$
has an extremum at $t=0$.  Indeed the derivative of $I$ is    $2( {x_1} (vx_1) + {y_1} (vy_1) +  {x_2} (vx_2 )  + {y_2} (vy_2 ) +  {x_3} (vx_3) + {y_3} (vy_3 ))$ which equals   $2(- vx_1 (0))  +2 (vx_2 (0))$ at $t=0$ for
solutions with an these initial positions.  Setting this quantity to zero,  together with the zero linear and angular momentum conditions,
yields all   velocity components in  terms of $v_x, v_y$ as above. 
The circular curve defining the points with zero energy (the $E=0$ curve) in the quadrant I of the $(v_x,v_y)-$plane
is defined by the function: $v_y=\sqrt{5/6-{v_x}^2}, v_x\in[0,\sqrt{5/6}], \sqrt{5/6}\approx 0.91287.$
As only negative energies (only bounded motions) have to be considered \cite{Suvakov:2014},
the 2D search domain is actually those bounded  by $v_x=0$ and $v_y=0$ axis and $E=0$ curve (see Fig. \ref{fig:search domain}).
If we denote the periods of the orbits with $T$, then the goal is to find triplets $(v_x, v_y, T)$
for which the periodicity condition $X(v_x,v_y,T)=X(v_x,v_y,0)$ is satisfied. For simplicity of notation, in what follows we will use the same notation for $v_x, v_y, T$ and their approximations and will not use indices for Newton's iterations, where they are not needed.

\subsection{Return proximity function}
\label{ss:return_proximity}

The function 
\begin{equation}
R(t)=\|X(v_x,v_y,t) - X(v_x,v_y,0)\|_{2}, \quad  t>0
\end{equation}
is called the return proximity function (the standard return proximity function).
The triplet $(v_x,v_y,T)$, $T>0$ corresponds to a periodic solution only if $R(T)=0.$
For approximate solution with approximations $(v_x,v_y,T)$, $R(T)$ is the measure of how close to a periodic solution we are.
$R(T)$ is the residual in which terms we define the convergence of the Newton's method for the standard periodicity condition.

\subsection{ Proximity function for the ``half period'' Eulerian condition} 
\label{ss:return_proximity}

As said in the introduction, the theoretical foundation of our new approach is
the ``half period'' property of the periodic orbits starting with Eulerian initial configuration.
The property says that all such solutions need to pass again through Eulerian configuration 
at the half period $T/2$ with the third body again in the middle. 
The proof of the property given by prof. Richard Montgomery at the University of California, Santa Cruz,
is in Appendix A of the paper. We want to mention that there is a small difference in the terminology we used and those in the proof. 
Namely the special initial conditions for the Eulerian configuration are called ``Eulerian half-twist conditions'',
but it is not a problem, as the proof should be regarded as an independent and self sufficient text.
Because of the particular choice of the center of mass of the initial configuration 
the third body at $T/2$ needs to be at $(0,0)$. If we introduce the vectors $X_1(t)={({x}_1, {y}_1, {vx}_1, {vy}_1)}^\top$,
$X_2(t)={(-{x}_2, -{y}_2, {vx}_2, {vy}_2)}^\top$, the ``half period'' Eulerian condition becomes: 
\begin{equation}
\label{halfperiod_cond}
X_1(v_x,v_y,T/2)=X_2(v_x,v_y,T/2)
\end{equation}
The proximity function for Eulerian condition is defined as: 
\begin{equation}
R_{e}(t)={\|X_1(v_x,v_y,t)-X_2(v_x,v_y,t)\|}_2, \quad t>0
\end{equation}
The triplet $(v_x,v_y,\overline{T})$, $\overline{T}>0$ corresponds to a solution of 
the Eulerian condition (equation) at $\overline{T}$ only if $R_{e}(\overline{T})=0.$
For approximate solution with approximations $(v_x,v_y,\overline{T})$, $R_{e}(\overline{T})$ is the measure of how close we are to an Eulerian configuration at $\overline{T}$. $R_{e}(\overline{T})$ is the residual in which terms we define the convergence of the Newton's method for the ``half period'' Eulerian condition. $\overline{T}$ is an approximation of $T/2$.

\section{Numerical methods}
\label{s:Numerical methods}

\subsection{Stages of the numerical search}
\label{ss:stages}

The numerical search for periodic orbits passing through Eulerian configuration consists of 3 stages:

Stage I: {\bf{Scanning stage:}} The candidates for correction (the initial approximations) 
are computed by scanning the 2D domain for the parameters $v_x,v_y$. This is the domain shown in Fig \ref{fig:search domain}. 
A quadratic grid is introduced in this domain. At each point $(v_x,v_y)$ of the grid, the system (\ref{equations}) 
with initial conditions (\ref{initcond}) is simulated up to some value
$T_0(v_x,v_y)/2$. The value $\overline{T}$ for which the minimum:
 
$$R_e(\overline{T})=\min\limits_{1<t<T_0/2}R_e(t)$$
is obtained, is computed. The triplet $(v_x, v_y, \overline{T})$ becomes a candidate 
for a periodic orbit if $R_{e}(\overline{T})$ is small and also is a local minimum on the grid.
The candidates are sorted in increasing with respect to $R_{e}(\overline{T})$ order
and then are processed in this order in the second stage.

Stage II: {\bf{Capturing stage:}} The candidates from stage I are corrected with a modification of Newton's method.
The modified method has a larger domain of convergence than the classic Newton's method.
The convergence of the method means that a solution is captured.
The method is applied with respect to the Eulerian condition and hence
the convergence is in terms of the corresponding from the first stage proximity function $R_{e}(\overline{T})$.

Stage III: {\bf{Verification stage:}} 
The captured from stage II solutions are computed with many correct digits 
with the classic Newton's method and the regular convergence of the method up to very high precision is checked.
Now the standard periodicity condition is always considered and the standard return proximity is used.

Although the modified Newton's method has a larger domain of convergence for solving a given equation than the classic method, 
its usage has a secondary importance in this paper. The main idea 
is to solve the Eulerian condition (equation) at $T/2$ instead of the equation from the standard periodicity condition. 
Not only the simulation time is divided by two. More importantly,
the properties of Newton's iterations change and the modified Newton's method with respect to the Eulerian 
condition has a much larger domain of convergence for a given periodic orbit in comparison 
to the same method with respect to the standard periodicity condition.

\subsection{Newton's method with respect to the standard periodicity condition}
\label{ss:Newton_standard}

As we mentioned before, we will not give additional notations for the approximations of the initial velocities and the period.
Let $(v_x,v_v, T)$ be such approximations, i.e. $X(v_x,v_y,T)\approx X(v_x,v_y,0).$
These approximations are improved with corrections $\Delta v_x,
\Delta v_y, \Delta T$ by expanding the periodicity condition:
$$X(v_x+\Delta v_x,v_y+\Delta v_y,T+\Delta T)= X(v_x+\Delta v_x,v_y+\Delta v_y,0)$$
in a multivariable linear approximation:

\begin{equation}
\begin{pmatrix}
x_1(T) \\
y_1(T) \\
x_2(T) \\
y_2(T) \\
x_3(T) \\
y_3(T) \\
{vx}_1(T)\\
{vy}_1(T)\\
{vx}_2(T)\\
{vy}_2(T)\\
{vx}_3(T)\\
{vy}_3(T)
\end{pmatrix}
+
\begin{pmatrix}
\frac{\partial x_1}{\partial v_x}(T) & \frac{\partial x_1}{\partial v_y}(T) & \dot{x}_1(T)\\
\frac{\partial y_1}{\partial v_x}(T) & \frac{\partial y_1}{\partial v_y}(T) & \dot{y}_1(T)\\
\frac{\partial x_2}{\partial v_x}(T) & \frac{\partial x_2}{\partial v_y}(T) & \dot{x}_2(T)\\
\frac{\partial y_2}{\partial v_x}(T) & \frac{\partial y_2}{\partial v_y}(T) & \dot{y}_2(T)\\
\frac{\partial x_3}{\partial v_x}(T) & \frac{\partial x_3}{\partial v_y}(T) & \dot{x}_3(T)\\
\frac{\partial y_3}{\partial v_x}(T) & \frac{\partial y_3}{\partial v_y}(T) & \dot{y}_3(T)\\
\frac{\partial {vx}_1}{\partial v_x}(T) & \frac{\partial {vx}_1}{\partial v_y}(T) & \dot{vx}_1(T)\\
\frac{\partial {vy}_1}{\partial v_x}(T) & \frac{\partial {vy}_1}{\partial v_y}(T) & \dot{vy}_1(T)\\
\frac{\partial {vx}_2}{\partial v_x}(T) & \frac{\partial {vx}_2}{\partial v_y}(T) & \dot{vx}_2(T)\\
\frac{\partial {vy}_2}{\partial v_x}(T) & \frac{\partial {vy}_2}{\partial v_y}(T) & \dot{vy}_2(T)\\
\frac{\partial {vx}_3}{\partial v_x}(T) & \frac{\partial {vx}_3}{\partial v_y}(T) & \dot{vx}_3(T)\\
\frac{\partial {vy}_3}{\partial v_x}(T) & \frac{\partial {vy}_3}{\partial v_y}(T) & \dot{vy}_3(T)
\end{pmatrix}
\begin{pmatrix}
\Delta v_x \\
\Delta v_y \\
\Delta T
\end{pmatrix}
=
\begin{pmatrix}
x_1(0) \\
y_1(0)  \\
x_2(0) \\
y_2(0) \\
x_3(0)  \\
y_3(0)  \\
{vx}_1(0) + \Delta v_x\\
{vy}_1(0) + \Delta v_y\\
{vx}_2(0) + \Delta v_x\\
{vy}_2(0) + \Delta v_y\\
{vx}_3(0)- 2 \Delta v_x\\
{vy}_3(0)- 2 \Delta v_y
\end{pmatrix}
\end{equation}
Finally, we obtain the following linear system with a $12\times 3$ matrix with respect to ${(\Delta v_x, \Delta v_y, \Delta T)}^\top$,
that have to be solved at each Newton's iteration.
\begin{equation}
\label{persystem}
\begin{pmatrix}
\frac{\partial x_1}{\partial v_x}(T) & \frac{\partial x_1}{\partial v_y}(T) & \dot{x}_1(T)\\
\frac{\partial y_1}{\partial v_x}(T) & \frac{\partial y_1}{\partial v_y}(T) & \dot{y}_1(T)\\
\frac{\partial x_2}{\partial v_x}(T) & \frac{\partial x_2}{\partial v_y}(T) & \dot{x}_2(T)\\
\frac{\partial y_2}{\partial v_x}(T) & \frac{\partial y_2}{\partial v_y}(T) & \dot{y}_2(T)\\
\frac{\partial x_3}{\partial v_x}(T) & \frac{\partial x_3}{\partial v_y}(T) & \dot{x}_3(T)\\
\frac{\partial y_3}{\partial v_x}(T) & \frac{\partial y_3}{\partial v_y}(T) & \dot{y}_3(T)\\
\frac{\partial {vx}_1}{\partial v_x}(T)-1 & \frac{\partial {vx}_1}{\partial v_y}(T) & \dot{vx}_1(T)\\
\frac{\partial {vy}_1}{\partial v_x}(T) & \frac{\partial {vy}_1}{\partial v_y}(T)-1 & \dot{vy}_1(T)\\
\frac{\partial {vx}_2}{\partial v_x}(T)-1 & \frac{\partial {vx}_2}{\partial v_y}(T) & \dot{vx}_2(T)\\
\frac{\partial {vy}_2}{\partial v_x}(T) & \frac{\partial {vy}_2}{\partial v_y}(T)-1 & \dot{vy}_2(T)\\
\frac{\partial {vx}_3}{\partial v_x}(T)+2 & \frac{\partial {vx}_3}{\partial v_y}(T) & \dot{vx}_3(T)\\
\frac{\partial {vy}_3}{\partial v_x}(T) & \frac{\partial {vy}_3}{\partial v_y}(T)+2 & \dot{vy}_3(T)
\end{pmatrix}
\begin{pmatrix}
\Delta v_x \\
\Delta v_y \\
\Delta T
\end{pmatrix}
=
\begin{pmatrix}
x_1(0) - x_1(T) \\
y_1(0) - y_1(T)  \\
x_2(0) - x_2(T) \\
y_2(0)-y_2(T) \\
x_3(0)-x_3(T)  \\
y_3(0) -y_3(T) \\
{vx}_1(0) -{vx}_1(T) \\
{vy}_1(0) -{vy}_1(T)\\
{vx}_2(0) -{vx}_2(T)\\
{vy}_2(0) -{vy}_2(T)\\
{vx}_3(0)-{vx}_3(T)\\
{vy}_3(0)-{vy}_3(T)\\
\end{pmatrix}
\end{equation}

\subsection{Newton's method with respect to the Eulerian condition}
\label{ss:Newton_euler}

Let $\overline{T}$ be an approximation of $T/2$ and the triplet
$(v_x,v_y,\overline{T})$ be an approximation of the ``half period'' Eulerian
condition at $T/2$, i.e. $X_1(v_x,v_y,\overline{T}) \approx X_2(v_x,v_y,\overline{T})$.
Then the Eulerian condition with corrections $\Delta v_x, \Delta v_y, \Delta \overline{T}$ is:
$$X_1(v_x+\Delta v_x, v_y+\Delta v_y, \overline{T}+ \Delta \overline{T})=X_2(v_x+\Delta v_x, v_y+\Delta v_y, \overline{T}+ \Delta \overline{T})$$
Expanding this equation in a multivariable linear approximation gives the following linear system with $4\times3$ matrices:

\begin{equation}
\begin{split}
\begin{pmatrix}
{x}_1(\overline{T})\\
{y}_1(\overline{T})\\
{vx}_1(\overline{T})\\
{vy}_1(\overline{T})
\end{pmatrix}
+
\begin{pmatrix}
\frac{\partial {x}_1}{\partial v_x}(\overline{T}) & \frac{\partial {x}_1}{\partial v_y}(\overline{T}) & \dot{x}_1(\overline{T})\\
\frac{\partial {y}_1}{\partial v_x}(\overline{T}) & \frac{\partial {y}_1}{\partial v_y}(\overline{T}) & \dot{y}_1(\overline{T})\\
\frac{\partial {vx}_1}{\partial v_x}(\overline{T}) & \frac{\partial {vx}_1}{\partial v_y}(\overline{T}) & \dot{vx}_1(\overline{T})\\
\frac{\partial {vy}_1}{\partial v_x}(\overline{T}) & \frac{\partial {vy}_1}{\partial v_y}(\overline{T}) & \dot{vy}_1(\overline{T})
\end{pmatrix}
\begin{pmatrix}
\Delta v_x \\
\Delta v_y \\
\Delta \overline{T}
\end{pmatrix}
=\\
\begin{pmatrix}
-{x}_2(\overline{T})\\
-{y}_2(\overline{T})\\
{vx}_2(\overline{T})\\
{vy}_2(\overline{T})
\end{pmatrix}
+
\begin{pmatrix}
-\frac{\partial {x}_2}{\partial v_x}(\overline{T}) & -\frac{\partial {x}_2}{\partial v_y}(\overline{T}) & -\dot{x}_2(\overline{T})\\
-\frac{\partial {y}_2}{\partial v_x}(\overline{T}) & -\frac{\partial {y}_2}{\partial v_y}(\overline{T}) & -\dot{y}_2(\overline{T})\\
\frac{\partial {vx}_2}{\partial v_x}(\overline{T}) & \frac{\partial {vx}_2}{\partial v_y}(\overline{T}) & \dot{vx}_2(\overline{T})\\
\frac{\partial {vy}_2}{\partial v_x}(\overline{T}) & \frac{\partial {vy}_2}{\partial v_y}(\overline{T}) & \dot{vy}_2(\overline{T})
\end{pmatrix}
\begin{pmatrix}
\Delta v_x \\
\Delta v_y \\
\Delta \overline{T}
\end{pmatrix}
\end{split}
\end{equation}
Finally, we obtain the following linear system with a $4\times 3$ matrix with respect to ${(\Delta v_x, \Delta v_y, \Delta \overline{T})}^\top$,
that have to be solved at each Newton's iteration.
\begin{equation}
\label{eulersystem}
\begin{split}
\begin{pmatrix}
\frac{\partial {x}_1}{\partial v_x}(\overline{T}) + \frac{\partial {x}_2}{\partial v_x}(\overline{T}) & \frac{\partial {x}_1}{\partial v_y}(\overline{T}) +
\frac{\partial {x}_2}{\partial v_y}(\overline{T}) & \dot{x}_1(\overline{T}) + \dot{x}_2(\overline{T})\\
\frac{\partial {y}_1}{\partial v_x}(\overline{T}) + \frac{\partial {y}_2}{\partial v_x}(\overline{T}) & \frac{\partial {y}_1}{\partial v_y}(\overline{T}) +
\frac{\partial {y}_2}{\partial v_y}(\overline{T}) & \dot{y}_1(\overline{T})+\dot{y}_2(\overline{T})\\
\frac{\partial {vx}_1}{\partial v_x}(\overline{T})-\frac{\partial {vx}_2}{\partial v_x}(\overline{T}) & \frac{\partial {vx}_1}{\partial v_y}(\overline{T})- 
\frac{\partial {vx}_2}{\partial v_y}(\overline{T})& \dot{vx}_1(\overline{T})-\dot{vx}_2(\overline{T})\\
\frac{\partial {vy}_1}{\partial v_x}(\overline{T})-\frac{\partial {vy}_2}{\partial v_x}(\overline{T}) & \frac{\partial {vy}_1}{\partial v_y}(\overline{T}) -
\frac{\partial {vy}_2}{\partial v_y}(\overline{T})& \dot{vy}_1(\overline{T})-\dot{vy}_2(\overline{T})
\end{pmatrix}
\begin{pmatrix}
\Delta v_x \\
\Delta v_y \\
\Delta \overline{T}
\end{pmatrix}
=
\begin{pmatrix}
-{x}_1(\overline{T})-{x}_2(\overline{T})\\
-{y}_1(\overline{T})-{y}_2(\overline{T})\\
{vx}_2(\overline{T})-{vx}_1(\overline{T})\\
{vy}_2(\overline{T}-{vy}_1(\overline{T})
\end{pmatrix}
\end{split}
\end{equation}
The linear systems (\ref{persystem}) and (\ref{eulersystem}) are solved in least square sense by QR-decomposition \cite{Demmel}.

\subsection{Modified Newton's method}
\label{ss:Modified_Newton}

Let the triplet $(v_x, v_y,\overline{T})$ be an approximate solution of the ``half period'' Eulerian condition (equation).
We correct and obtain the next approximation with the classic Newton's method this way:
$$v_x := v_x + \Delta v_x,\hspace{0.2 cm} v_y := v_y + \Delta v_y, \hspace{0.2 cm} \overline{T} := \overline{T} + \Delta \overline{T}$$
At stage II of the numerical search (when capturing the periodic orbits)
we use a modification of Newton's method based on the continuous  analog of Newton's method  \cite{CANM}. We introduce the parameter $p_k$:
$0<p_k\leq1$, where $k$ is the number of the iteration.
Now we correct this way:
$$v_x := v_x + p_k\Delta v_x, \hspace{0.2 cm} v_y := v_y + p_k\Delta v_y, \hspace{0.2 cm} \overline{T} :=\overline{T} + p_k\Delta \overline{T}$$
Let $R_k$ be the value of the proximity function $R_e( \overline{T})$ (the residual) at the $k$-th iteration.
With a given $p_0$ the next $p_k, k=1,2,...$ is computed with the following adaptive algorithm \cite{CANM}:

\begin{equation}
p_k = \left \{
               \begin{array}{ll}
                \min(1,\ p_{k-1} R_{k-1} / R_k), &
                 R_k \leq R_{k-1}, \\\\
                \max(p_0,\ p_{k-1} R_{k-1} / R_k), &
                 R_k > R_{k-1},
               \end{array}
                                  \right .
\end{equation}
The modification of  Newton's method does not add any technical difficulties, because we have to solve the same linear system.
The important thing is that for a given equation that have to be solved, this method has a larger domain of convergence and makes it possible to find
more periodic orbits for a given search grid. When $p_k=1$ for all $k$, the method matches with the classic Newton's method. We usually take  $p_0=0.2$. As our numerical experience shows, it is usually more efficient to use the modified Newton's method
instead of obtaining a similar result with a finer search grid and the classic Newton's method.

\subsection{Computing the elements of the matrices}
\label{ss:elements}

To compute the elements of the  matrices for the linear systems (\ref{persystem}) and (\ref{eulersystem})
we have to add to the system (\ref{equations}) the 24 differential  equations for the
partial derivatives with respect to the parameters $v_x, v_y$:
$$\frac{\partial x_i}{\partial v_x}(t),\frac{\partial y_i}{\partial v_x}(t),
\frac{\partial {vx}_i}{\partial v_x}(t),\frac{\partial {vy}_i}{\partial v_x}(t),
\frac{\partial x_i}{\partial v_y}(t),\frac{\partial y_i}{\partial v_y}(t),
\frac{\partial {vx}_i}{\partial v_y}(t),\frac{\partial {vy}_i}{\partial v_y}(t), i=1,2,3.$$
These equations can be obtained by differentiation of the system  (\ref{equations}) with respect to the parameters $v_x, v_y$,
but we do not need them in explicit form. We need, however, the initial conditions. They are:
$$\frac{\partial x_i}{\partial v_x}(0)=\frac{\partial y_i}{\partial v_x}(0)=\frac{\partial x_i}{\partial v_y}(0)=\frac{\partial y_i}{\partial v_y}(0)=0, i=1,2,3$$
$$\frac{\partial {vy}_i}{\partial v_x}(0)=0, i=1,2,3, \frac{\partial {vx}_1}{\partial v_x}(0)=\frac{\partial {vx}_2}{\partial v_x}(0)=1, \frac{\partial {vx}_3}{\partial v_x}(0)=-2$$

$$\frac{\partial {vx}_i}{\partial v_y}(0)=0, i=1,2,3, \frac{\partial {vy}_1}{\partial v_y}(0)=\frac{\partial {vy}_2}{\partial v_y}(0)=1, \frac{\partial {vy}_3}{\partial v_y}(0)=-2$$
Although the system (\ref{eulersystem}) use explicitly only a part the partial derivatives (those for body 1 and body 2), 
all 24 partial derivatives have to be computed, because the differential equations for them are coupled.
At each step of Newton's or modified Newton's method we have to solve a system of 36 ODEs in the interval $[0,T]$ or $[0,\overline{T}]$ with initial conditions corresponding to $v_x, v_y$. 12 ODEs come from system (\ref{equations}) and 
24 ODEs come from the equations for the partial derivatives with respect to $v_x, v_y.$

A crucial decision for the success of finding periodic orbits is the choice of the numerical algorithm for solving 
the original system  (\ref{equations}) for stage I and the system of 36 ODEs for stage II and stage III.
To follow the trajectories accurately for a sufficiently long time,
we need to overcome the sensitive dependence on the initial conditions and to handle the close encounters of the bodies accurately.
To do this, we used (as in \cite{Liao:2017}) high order high precision Taylor series method with adaptive stepsize control \cite{Jorba, Barrio:2006, Barrio:2011, Izzo}.

\subsection{Taylor series method and stepsize control}
\label{TSM}

For the initial value problem 
$\dot{u}(t)=f(u,t), ~ u_0=u(0),$ 
the N-th order Taylor series method for finding an approximate solution $U(t) \approx u(t)$ is given by:
$$U(t+\tau)=\sum_{i=0}^{N} U^{[i]}\tau^i, \hspace{0.3 cm} U^{[i]}=\frac{1}{i!}\frac{dU^{(i)}(t)}{dt^i},$$
where the coefficients $U^{[i]}$ are called normalized derivatives.
With already computed normalized derivatives, the series is evaluated by Horner's rule.

The use of an adaptive time stepsize strategy is crucial
for the three-body problem, especially for handling the close encounters of the bodies. 
The time stepsize  $\tau$ is determined
according to the paper \cite{Jorba}. Setting the absolute and the relative tolerances from this paper to be equal, we
obtain the following formula for $\tau$:
\begin{equation}
\label{stepsize}
\tau = \frac{e^{-0.7/(N-1)}}{e^{2}} \min\left\{   {\left(\frac{1}{{\|\mathbf{U^{[N-1]}} \|}_{\infty}}\right)}^{\frac{1}{N-1}}, {\left(\frac{1}{{\|\mathbf{U^{[N]}}\|}}_{\infty}\right)}^{\frac{1}{N}}\right\}
\end{equation}
The time stepsize is determined by the last two terms of the Taylor expansions. Terms are two in order to avoid problems with odd and even functions.
The expression in brackets is essentially an estimation of the smallest radius of convergence among Taylor series
of all the scalar variables accordingly to the Cauchy-Hadamard theorem.

A fixed order Taylor series method is used. The order depends on the set precision.
In practice reducing the discretization (truncation) error by increasing the order $N$
stops at some point due to the rounding error.
As numerical experiments show, order 22 per each 64 bit mantissa (bit of precision) is an optimal choice.
Let us note that for this choice of the order our formula (\ref{stepsize}) 
for the time stepsize gives almost the same value as the formula (25)
for the optimal stepsize in paper \cite{Hu}.

The most technical part of the Taylor series method is the computation of the normalized derivatives. 
They are computed by the rules of automatic differentiation \cite{Jorba, Barrio:2006, Barrio:2011}. 
The detail formulas for the normalized derivatives, 
including those for the partial derivatives with respect to $v_x, v_y$,  can be seen in the work \cite{Hristov:2021}.

\section{Numerical results}
\label{s:Numerical results}

\subsection{Parameters of simulations and accuracy analysis}
\label{ss:Accuracy analysis}

The ODE solver based on high order and high precision Taylor series method (TSM) is the main building block of our numerical search.
If we assume its correctness, it is not difficult to check the accuracy of Newton's method by testing
its convergence in terms of the return proximity. As a benchmark for our ODE solver, we used the high precision numerical results in the work \cite{benchmark}. All given digits in the numerical results in this work are the same as ours.

At stage I (Scanning stage) we introduce a quadratic grid with size $3*2^{-14} \approx 1.831*10^{-4}$  in the domain 
in Figure \ref{fig:search domain}. This grid size corresponds to about 19 million points in the domain. We search for periodic orbits with scale invariant periods $T^{*}=T|E(v_x,v_y)|^{3/2}<70.$
We handle the singularity when approaching the boundary curve $E=0$ by limiting the periods up to some relatively large number. 
We choose this number to be 1000. In each point of the grid $(v_x,v_y)$, we consider periods less than $T_0/2$, where $T_0(v_x,v_y)=\min(1000,70/|E(v_x,v_y)|^{3/2})$,
i.e. the smallest of the number 1000 and the number that gives the scale-invariant period $T^{*}=70$.
In other words, between the curves $E=-(70/1000)^{2/3}$ and $E=0$ our search is limited to periods up to 1000.
We simulate the system (\ref{equations}) with initial conditions (\ref{initcond}) up to $T_0/2$ with precision of 128 bit ($\approx 38.5$ decimal digits) and 44 order TSM. As mentioned in subsection \ref{ss:stages}, 
at each grid point $(v_x,v_y)$ we compute the time $\overline{T}$ at which the minimum:

$$R_e(\overline{T})=\min\limits_{1<t<T_0/2}R_e(t)$$
is obtained. The accuracy of $\overline{T}$ and $R_e(\overline{T})$ may be not good enough, if we compute $R_{e}(t)$ within the time stepsize of TSM.
To improve the accuracy, we use the property of TSM to provide a dense output within the step with a small additional cost.
With already computed derivatives at point $t$ and determined step $\tau$, it is straightforward
to compute a high precision approximation at any point in $(t, t+\tau)$ with Horner's rule.
When $R_{e}(t)$ is small, we divide the interval $(t, t+\tau)$ by 1000 and compute $R_{e}$ at each point.
Actually, this way of  providing a dense output is similar to the interpolation process commonly used with other integration methods.
Candidates for the second stage become those triplets $(v_x,v_y, \overline{T})$ for which $R_e(\overline{T})$ is
small $(<0.3)$ and also $R_e(\overline{T})$ is a local minimum on the search grid.

At Stage II (Capturing stage) we use the  modification of Newton's method 
starting with initial approximations from stage (I).
The system (\ref{eulersystem}) is solved at each iteration, i.e. the Eulerian condition is considered. The convergence is in terms of the corresponding from the first stage proximity function $R_{e}(\overline{T})$. Precision of 192 bit ($\approx 57.8$ decimal digits) and 66 order TSM is used. A periodic solution is captured if $R_{e}(\overline{T})<10^{-20}$. Each captured solution is additionally specified up to $R_{e}(\overline{T})<10^{-50}$  
by computations with increased precision of 320 bit ($\approx 96.3$ decimal digits) and 110 order TSM.

At Stage III (Verification stage) we increase the precision and the order of the method.
At this stage we use the classic Newton's method with respect to the standard periodicity condition 
and the convergence is in terms of the standard return proximity $R(T)$.
Now the iterations are until convergence (until the return proximity $R(T)$ stops to decrease).
In three substages we gradually increase the order and precision and use the approximations obtained from the previous substage.
The minimal return proximities (the return proximity at which the convergence stops) differ for different initial conditions. 
These minimal values generally depend on:

1) The errors for the computed solution and elements 
of the matrix in system (\ref{persystem})
accumulated for one period of integration with TSM. 
In turn, these errors depend on the maximal Lyapunov exponent (for unstable solutions)
and on the level at which the discretization (truncation) error of TSM is decreased.

2) The error for solving the least square problem with QR-decomposition.

How the discretization error of TSM and the error for solving the least square problem depend on the closeness of encounters is an important, 
but also a difficult question, which answer we postpone for the future.
From the numerical experience, however, we can conclude that we can always
successfully simulate a given solution with close encounters, by increasing the precision and order of TSM.

At the first substage we use precision of 448 bit ($\approx134.9$ decimal digits) and 154 order TSM.
The minimal return proximities for all i.c.s. are approximately in the interval $(6.50*10^{-134}, 1.29*10^{-109}).$
At the second substage the precision is 576 bit ($\approx173.4$ decimal digits), the order - 198
and the corresponding interval of minimal return proximities is $(2.53*10^{-172}, 3.09*10^{-148}).$
At the third substage the precision is 704 bit ($\approx211.9$ decimal digits), the order - 242
and the interval is $(8.26*10^{-211}, 1.10*10^{-186}).$ 
The minimal return proximity is closed to the used precision for the stable periodic orbits.
Its maximal value is approximately 26 digits below the used precision for all three substages
and is obtained for the same initial condition.
These results are in consistency with the results for the comparison of the obtained first digits for the three substages.
By comparing the obtained digits for the i.c.s and periods $T$ at the three substages, we found that more than first 100 digits are the same.
We also made an additional verification by dividing by four the prescribed from TSM time stepsize at the first substage and obtain the same first 100 digits.
By comparing the digits for the second and third substage, we see that more than 140 of them are the same. 
So, we can safely assume that for the last most precision computations 180 digits are correct. 

It should be noted that for collisionless orbits
the convergence of Newton's method has the property that
the number of time steps at each iteration becomes fixed once we are sufficiently
close to a periodic solution. This property is also checked for all i.c.s and for all three substages.
Furthermore, the convergence in terms of the return proximity agrees very well
with the theoretical quadratic convergence of Newton's method.

In addition, we investigate the linear stability of the found orbits. 
Without going into too much detail, we will briefly describe these computations.
For the stability study we use the most accurate (with 180 correct digits) initial conditions and periods.
The linear stability of an orbit is investigated by computing the eigenvalues of its monodromy matrix \cite{Hristov:Bor22}.
The elements of the monodromy matrix are computed in the same way as the partial derivatives in the systems (\ref{persystem}) and (\ref{eulersystem}) -
with TSM by using the rules of automatic differentiation.
Two computations are conducted, one with  precision of 448 bit ($\approx134.9$ decimal digits) 
and 154 order TSM and second with precision of 576 bit ($\approx173.4$ decimal digits) and 198 order TSM.
As more than first 100 digits for all monodromy matrices are the same, 
we can safely assume that the first 130 digits for the second computations are correct.
The Multiprecision Computing Toolbox \cite{Advanpix} for MATLAB\textsuperscript \textregistered \cite{Matlab}
is used for computing the eigenvalues.
Four of the eigenvalues determine the stability of an orbit \cite{Hristov:Bor22}.
More than the first 30 digits of these four eigenvalues are the same for
the two conducted with the toolbox computations - with 70 and 130 decimal digits of precision. 

One should not stay however with the impression, that all given from us 100
significant digits are necessary to make a good simulation for the found periodic orbits.
For all of them quadruple precision ($\approx 35$ digits) and good ODE solver are enough to close very well one period.
More computed correct digits are given to show that in principle these computations can be done
(that the Newton's method has regular convergence). Obtaining the solutions with many
correct digits and the regular high precision convergence of Newton's method  give us a much
greater confidence (rigor) about the existence of the solutions we discovered numerically.

The extensive high precision computations are performed in Nestum cluster, Sofia Tech Park, Bulgaria \cite{Nestum},
where the GMP library (GNU multiple precision library) \cite{GMP} is installed.

\subsection{The initial conditions found}
\label{ss:Initial}

Let us recall that the two types of symmetries on the shape sphere established in \cite{Suvakov:2013} 
correspond to different properties of the Eulerian configuration at $t=T/2$. 
For symmetry of type (I) the distance between the bodies at $t=T/2$ is 1 - the same as the distance at $t=0.$ 
In this case the Eulerian configuration at $t=T/2$ is essentially the same.
For symmetry of type (II) the distance between the bodies at $t=T/2$ is different from 1 
and the Eulerian configuration is essentially different from that at $t=0$.
The second i.c. is obtained by rescaling the positions and velocities at $t=T/2$.  
So, the solutions of type (I) are presented by one i.c. (one point in the search domain),
but those from type (II) - by two i.c.s (two points in the search domain).

An interesting observation is that the lines  defined by the Eulerian configurations at $t=0$ and $t=T/2$ are generally different.
An exception is a special case of solutions of type (I) for which the line is the same. 
More precisely, in this special case the  positions are exactly the same, but the velocities are opposite. 
These special periodic orbits turned out to be in addition free-fall orbits \cite{Montgomery:2023, Liao:2019, Freefall:2024}.
They correspond to the orbits with a central symmetry in the real space that are described in \cite{Montgomery:2023}. 
For them the bodies at $t=T/4$ and $t=3T/4$ are stopped (with zero velocities).
A small number (35) of this kind of orbits are found. Some of them are previously found in \cite{Freefall:2024}
(by a basically different searching approach).
 
We want to mention that the connection between the symmetries on the shape sphere and the Eulerian configuration at $t=T/2$
is checked by our very high accuracy results, but it needs of course a strict mathematical proof.
The statement that the line on which the bodies lie at $t=T/2$ is always different from those at $t=0$,
except for the special free-fall case, also needs a proof. 
Some connection might also be expected between the Eulerian configuration at $t=T/2$ and the
algebraic exchange symmetries of the free group elements established in \cite{Suvakov:2013}.

9584 different i.c.s for periodic collisionless orbits were captured at stage II of the numerical search.
This is the number of i.c.s obtained after a test for multiples of the fundamental periods.
6502 of i.c.s are in couples with the same $T^*$, thus representing 3251 solutions with 
symmetry of type (II). We simulated the other 3082 single i.c.s up to $t=T/2$ and analyzed the Eulerian configuration at $t=T/2$.
We found that 2847 of them are actually of symmetry type (II), but the second i.c. was missed. 
Adding these 2847 i.c.s we finally obtained 12431 i.c.s corresponding  to 6333 different solutions, 
6098 of type (II) and only 235 of type (I). 
35 of the orbits of type (I) are representatives of the mentioned above special free-fall orbits. 
Let us mention that the method does not exclude capturing of orbits with $T^*$, which is a little larger than 70. 
We have not removed these orbits from our list.
One can download the data $(v_x, v_y, T, T^*)$ with  100 correct digits for the found i.c.s from \cite{site}.
The i.c.s are ordered by $T^*$.

It is not easy to count exactly how many of the found i.c.s are new, 
because of the many numerical searches conducted by different authors in the last decade. 
Since only 33 of the i.c.s in the large database of 695 i.c.s from \cite{Liao:2017} are with $T^* < 70$,
we can confidently assume that most of the found i.c.s (say $>10000$) are new.
Of course  all 33 i.c.s with $T^*<70$ from the database from \cite{Liao:2017} were captured successfully from us.
In fact, the large number of new i.c.s is a consequence of the efficiency of our new approach,
but is not the primary goal of the paper. Actually this numerical search can be regarded as a test one,
because a relatively coarse search grid and relatively short scale invariant periods are considered. 
We expect that choosing a finer search grid and assuming the same condition $T^*<70$ will produce much more i.c.s. 
Also the number of orbits is expected to grow rapidly by increasing $T^*$.

\begin{figure}
\centerline{\includegraphics[width=0.8\columnwidth,,keepaspectratio]{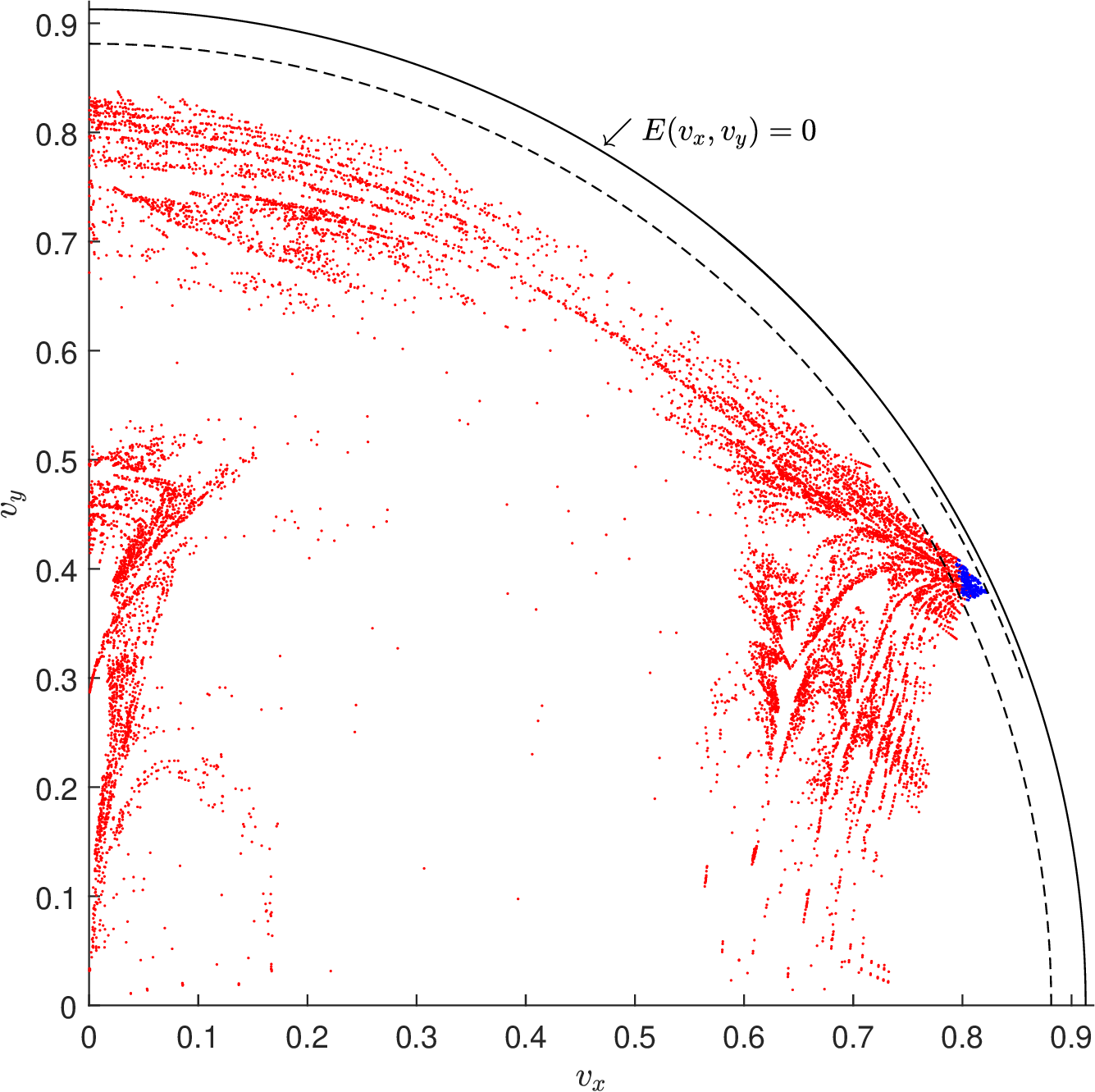}}
\caption{Distribution of the found i.c.s in the search domain}
\label{Distribution}
\end{figure}

\begin{figure}
\centerline{\includegraphics[scale=0.53]{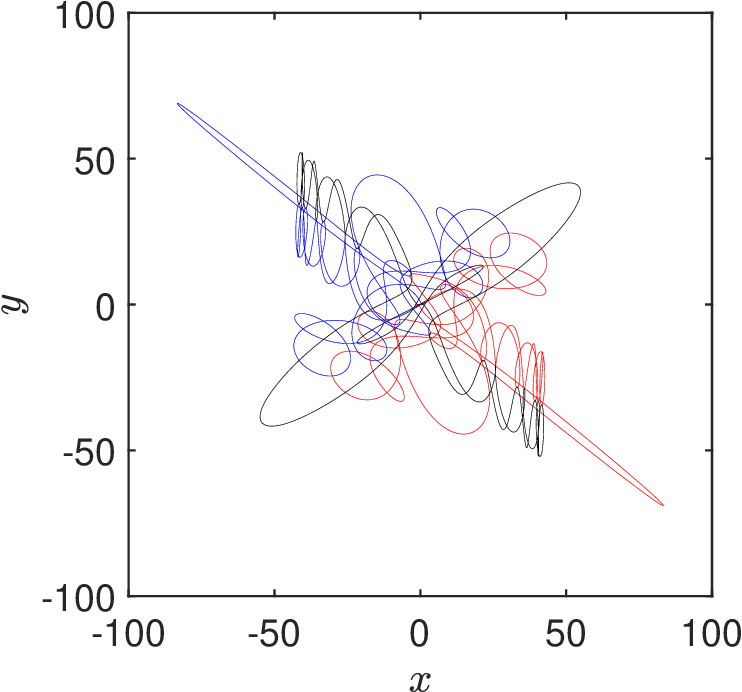}
\includegraphics[scale=0.53]{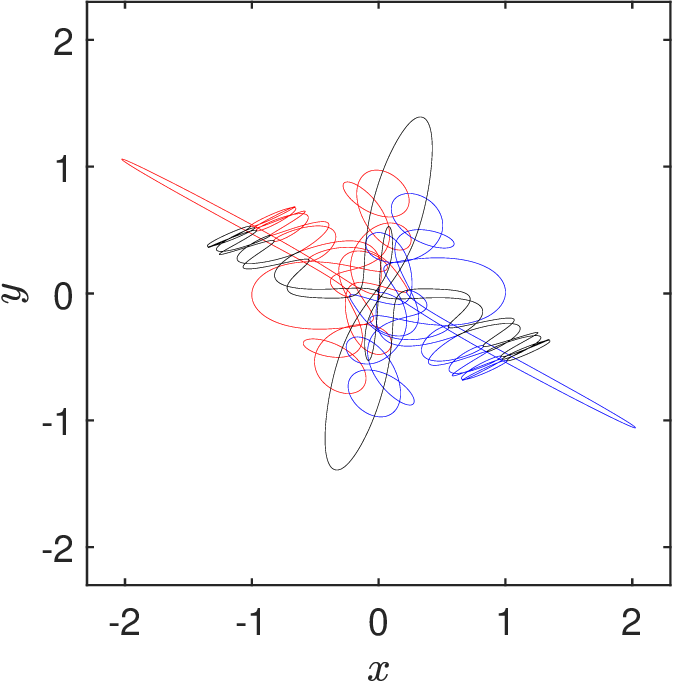}}
\vspace{0.5 cm}
\caption{\small{Real space plots for: Left -- for i.c. 1 (see Table 1);  and Right -- for  i.c. 2: (see Table 1)}}
\label{fig:ic1ic2}
\end{figure}

The distribution of the i.c.s in the search domain is shown with red points in Figure \ref{Distribution}.
We expect that the empty regions in the search domain will be more and more filled with points, when $T^*$ is increased.
It can be seen that the qualitative picture is quite different from that in the analogous figure in \cite{Liao:2017}.
So, one needs to be very careful when drawing conclusions from such figures and always has
to take into account the limitations on the considered periods and of course the efficiency of the used method.
Our choice to treat the singularity when approaching the boundary $E=0$ 
seems to be sufficient, as the red points cross the curve
$E=-(70/1000)^{2/3}$ (the long dashed line in Figure \ref{Distribution}) in a small region. 
Actually many of the points behind the long dashed line are not obtained directly, but are
secondary i.c.s of solutions of symmetry type (II).
To test the hypothesis for the existence of a sequence of periodic orbits 
that approaches the $E=0$ curve with ever increasing periods, we conducted 
an additional search in the region, where the red points cross the long dashed line.
We considered a new condition for the periods: $T_0(v_x,v_y)=\min(10000,70/|E(v_x,v_y)|^{3/2})$
and two times finer search grid. The short dash line in Figure \ref{Distribution} is the curve $E=-(70/10000)^{2/3}$.
The results support the hypothesis, 332 additional i.c.s (the blue points) were found which are even closer to the 
curve $E=0$. These 332 additional i.c.s with 100 correct digits can also be downloaded from \cite{site}.
To get even more closer to the boundary, one should consider longer periods, but also a much finer search grid.
The longest period for these 332 orbits is $\approx 9057$. The scale invariant period for the longest orbit is a little greater
than 70, but as mentioned before, the method does not exclude to find such orbits.
This solution is of symmetry type (II). 
Its secondary i.c. is close to the $v_y$-axis of the search domain and is with energy with a larger absolute value. 
35 digits data for the two i.c.s representing the same solution is given in Table 1.
Their real space plots  are shown in Figure \ref{fig:ic1ic2}.

\begin{table}
 \begin{tabular}{ p{0.1cm} p{6cm} p{6cm}}
 \hline
 $i.c.$ & $\hspace{2cm}v_x$ & $\hspace{2cm}v_y$\\
 {} & $\hspace{2cm}T$ & $\hspace{2cm}T^*$\\
  \hline
 1 &  0.82272349245737688870869449736746568e0 & 0.37823547230900304229962333317421811e0 \\
 {} & 0.90567760883687670170474569285377982e4 & 0.72976361086133980846350413612442343e2 \\
 \hline
 2 &  0.28923162959044843502097731887636809e-1 & 0.44566664996474142695785275860859663e0 \\
 {} & 0.27828627114705281918009819631639055e2 & 0.72976361086133980846350413612442343e2 \\
  \hline
\end{tabular}
{\caption {35 correct digits for the i.c. with the longest period and its secondary i.c.}}
\label{tab:table1}
\end{table}

The additional computations for linear stability shows that only 8 i.c.s (corresponding to 7 different solutions) are stable.
All stable solutions are old ones. The 7 stable solutions are the famous figure-8 orbit \cite{Moor, Figure8}, 3 solutions from the table in \cite{Suvakov:2013}, 2 solutions from \cite{Liao:2017} and the first known different from figure-8 stable choreography \cite{Suvakov:2014b}.
All the stable solutions are of symmetry type (I), except of the choreography from \cite{Suvakov:2014b} 
which is of symmetry type (II) and is presented by two i.c.s.
The fact that we did not find new periodic orbits for the considered small scale invariant periods is expected, 
because stable periodic orbits are much easier to be found numerically and many of them are already found in the previous searches.
The four eigenvalues which determine the linear stability \cite{Hristov:Bor22} for all 12,431 i.c.s are given  with 20 correct digits in \cite{site}.

The real space plots of the first 200 i.c.s in the list, the 35 special free-fall orbits, 
and the old 7 stable orbits can also be downloaded from \cite{site}.
The 100 correct digits data for the old 7 stable orbits and the free-fall orbits are in separate files there.
At last, one can find in \cite{site} the syzygy sequences of all i.c.s computed by Tanikawa and Mikkola's syzygy counting method \cite{Tanikawa:2008,Tanikawa:2015}. Corresponding to the usual bodies's numeration at the initial configuration, the sequences start and finish with symbol 2, as the initial and the final syzygies (at $t=0$ and $t=T$) are counted as different.

\section{Concluding remarks and Outlook}
\label{s:Conclusions}
$_{}$

1) A new  efficient searching approach for periodic three-body equal-mass orbits passing through Eulerian configuration is proposed.
The approach is based on the property of these orbits to pass again trough  Eulerian configuration at the half period $T/2$.
The results of the conducted numerical search clearly demonstrate  the efficiency of the approach. 12,431 i.c.s for periodic collisionless orbits
(most of which new ones) are found. The results are presented as a high precision (100 significant digits) database.

2) A connection between the properties of the Eulerian configuration at $T/2$ and the previously known two types of symmetries
on the shape sphere is established. Depending on the type of symmetry the periodic orbits are presented by one or two i.c.s in the
search domain. This connection of course needs a further mathematical proof. 

3) A small number (35) of special type of orbits that pass trough both Eulerian and free-fall configurations are observed.
It is a conjecture that this kind of orbits is the only one for which the lines for Eulerian configuration at $t=0$ and $t=T/2$ are the same.
   
4) The results suggest the existence of a sequence of periodic orbits that approaches the boundary of the search domain $E=0$
with ever increasing periods. Additional numerical check with longer periods and finer search grid in this region is needed.
   
5) Stability of orbits is also investigated. No new stable orbits are found.
The four eigenvalues which determine the linear stability are computed for all i.c.s and given with 20 correct digits.
This large high precision eigenvalues's database can be useful for future studies.


\bmhead{Acknowledgments}
I.H. thanks Richard Montgomery for the prompt reply of the questions concerning the ``half period'' property and for the proof of this property.
I.H. also thanks  Marija Jankovi\'c and Kiyotaka Tanikawa for advices and answers to theoretical questions.
The authors acknowledge the access to the Nestum cluster at HPC Laboratory, Research and Development and Innovation Consortium, Sofia Tech Park. The work of I.H. is financed by the European Union-NextGenerationEU, through the National Recovery and Resilience Plan of the Republic of Bulgaria, project number BG-RRP-2.004-0008-C01. The work of R.H. is supported by the EuroHPC JU through the project EuroCC2, Grant Agreement number 101101903, and by
the Ministry of Education and Science of Bulgaria through the Grant Agreement D01-168/28.07.2022.

\appendix

\section{Appendix: The proof of the ``half period'' property (by Richard Montgomery)}
Use the notation of the body of the text so
that positions are $r_i = (x_i , y_i)$
and velocities are $v_i = (vx _i, vy _i)$, both of which are two-vectors. 
We assume the three masses are equal: $m_1 = m_2 = m_3$  and that
the center of mass and linear momentum are both zero:
$r_1 + r_2 + r_3 = 0 = v_1 + v_2 + v_3$.

An initial condition is given by specifying initial positions  $r = (r_1, r_2, r_3)$
and initial velocities $v = (v_1, v_2, v_3)$ at  some time $t =t_*$.
We  call the initial configuration $r$ Eulerian (relative to the choice 1-2 from $\{1,2,3 \}$)
if $r_1 = -r_2$.  An Eulerian configuration necessarily has $r_3 = 0$.
We call the full set of  initial conditions   ``Eulerian half-twist'' at time $t = t_*$
if the configuration is Eulerian, so that   $r_1 (t_*)  = -r_2 (t_*)$ and
if the velocities satisfy  $v_1 (t_*) = v_2 (t_*)$. 
Necessarily we have that   $r_3 (t_*)  =0$ and $v_3 (t_*)  = -2 v_1 (t_*)$.

\begin{proposition} 
If a  three-body solution $r(t)$
has  Eulerian half-twist initial conditions at time $t =0$
and is periodic of period $T$ then it has  Eulerian
half-twist initial conditions at the half period, time $t = T/2$.
\end{proposition}

{\bf Proof.}  
Consider Eulerian half-twist initial conditions
$r(0) = (r_1 (0), -r_2 (0),  0)$ and $v(0) = (v_1 (0), v_2 (0), -2 v_1 (0))$
and write   $r(t) = (r_1 (t), r_2 (t), r_3 (t)$ for the corresponding solution
having these initial conditions at time $t = 0$.  
Now if $q(t) = (q_1 (t), q_2 (t), q_3 (t))$ is a solution to the equal mass 3 body problem, so is the curve 
$(-q_2 (-t), -q_1 (-t),  -q_3 (-t))$.  Thus $(-r_2 (-t),  -r_1 (-t), -r_3 (-t))$
is also a solution.  But the initial conditions of this second curve are identical
to those of $r(t)$ at $t = 0$.   Since the two curves have the same
initial conditions and satisfy the same differential equations they are in fact the same curve: 
\begin{equation} 
r_1 (t) = -r_2 (-t),  r_2 (t) = -r_1 (-t),  r_3 (t) = -r_3 (-t), 
\label{Isosc2} 
\end{equation} 
valid for all times $t$.
We will call this the {\it symmetry condition}. 

Now suppose our solution  curve $r(t)$ is periodic of period $T$:  $r(t + T) = r(t)$.
Taking $t = -T/2$ we get $r(T/2) = r(-T/2)$ or
$$r_1 (T/2) = r_1 (-T/2),  r_2 (T/2) = r_2 (-T/2),  r_3 (T/2) = r_3 (-T/2)$$
But by the   symmetry condition (\ref{Isosc2}) we then have that
$r_3 (T/2) = 0$ and $r_2 (T/2) = -r_1 (T/2)$ - that is, the configuration is isosceles at time $t = T/2$.

To verify the Eulerian half-twist  velocity conditions at $t = T/2$,  apply the symmetry condition
(\ref{Isosc2}) for $t = T/2 + h$ and combine it with the periodicity condition.
The symmetry condition yields that  
$$(r_1 (-T/2 - h), r_2 (-T/2 -h), r_3 (-T/2 -h)) = (-r_2 (T/2 + h),  -r_1 (T/2 + h), - r_3 (T/2 +h)).$$
Since $-T/2 -h + T = T/2 -h$  the periodicity condition is that  
$$r_1 (-T/2 - h) = r_1 (T/2 -h),
r_2 (-T/2 - h) = r_2 (T/2 -h), r_3 (-T/2 - h) = r_3 (T/2 -h)$$
 so combining the two conditions yields  
 $$(r_1 (T/2 - h), r_2 (T/2 -h), r_3 (T/2 -h)) = (-r_2 (T/2 + h),  -r_2 (T/2 + h), - r_3 (T/2 +h))$$
Differentiate this identity at $h = 0$ to arrive at
$-v_1 (T/2) = -v_2 (T/2)$ or $v_1 (T/2) = v_2 (T/2)$.
(The  zero linear momentum condition guarantees that  $v_3 (T/2) = -2 v_1 (T/2)$.)
We have established that at the half-period the solution has Eulerian half-twist initial conditions.  QED

%\end{comment}

\end{document}